\documentclass[aps,prl,twocolumn, english, floatfix, 
superscriptaddress, 10pt]{revtex4-1}
\usepackage[T1]{fontenc}
\usepackage[latin1]{inputenc}
\usepackage{lmodern}
\expandafter\let\csname equation*\endcsname\relax
\expandafter\let\csname endequation*\endcsname\relax
\usepackage{amsmath}
\usepackage{txfonts}
\usepackage{esint}
\usepackage{amssymb}
\usepackage{graphicx}
\usepackage{xcolor}
\usepackage{natbib}
\usepackage{bm}
\usepackage{bbold}
\usepackage[cal=boondoxo]{mathalfa}
\usepackage{comment}

\graphicspath{{./img/}}

\def\pM{\mathrel{\raise 2pt \hbox{\tiny(}\!\raise 1pt \hbox{+}\settowidth {\dimen03} {+}\hskip-\dimen03 \raise -2.4pt \hbox {$-$} \!\raise 2pt \hbox{\tiny)}}}

\renewcommand*{\vec}[1]{{\boldsymbol{\mathrm{#1}}}}

\usepackage[unicode=true,pdfusetitle,
bookmarks=true,bookmarksnumbered=true,bookmarksopen=true,bookmarksopenlevel=2,
breaklinks=false,pdfborder={0 0 0},backref=false,colorlinks=true]{hyperref}
\hypersetup{colorlinks=true, linkcolor=blue, citecolor=blue, urlcolor=blue}

\usepackage{bookmark}
\newcommand*{\heading}[1]{\belowpdfbookmark{#1}{#1}{\bfseries\textit{#1.---}}\ignorespaces}
\let \section \heading

\begin{document}

\title{Flat and tunable moir\'e phonons in twisted transition-metal dichalcogenides}


\newcommand{\Harvard}{John A.~Paulson School of Engineering and Applied Sciences, Harvard University, Cambridge, Massachusetts 02138, USA}
\newcommand{\Columbia}{Department of Physics, Columbia University, New York, NY 10027, USA}
\newcommand{\Minnesota}{School of Physics and Astronomy, University of Minnesota, Minneapolis, MN 55455, USA}
\newcommand{\Yale}{Department of Physics, Yale University, New Haven, Connecticut 06520, USA}



\author{Alejandro Ramos-Alonso}
\affiliation{\Columbia}

\author{Benjamin Remez}
\affiliation{\Yale}

\author{Daniel Bennett}
\affiliation{\Harvard}

\author{Rafael M.~Fernandes}
\email[]{rfernand@umn.edu}
\affiliation{\Minnesota}

\author{H{\'e}ctor Ochoa}
\email[]{ho2273@columbia.edu}
\affiliation{\Columbia}

\begin{abstract}
Displacement fields are one of the main tuning knobs employed to engineer flat electronic band dispersions in twisted van der Waals multilayers.
Here, we show that electric fields can also be used to tune the phonon dispersion of moir\'e superlattices formed by non-centrosymmetric materials, focusing on twisted transition metal dichalcogenide homobilayers. 
This effect arises from the intertwining between the local stacking configuration and the formation of polar domains within the moir\'e supercell.
For small twist angles, increasing the electric field leads to a universal moir\'e phonon spectrum characterized by a substantially softened longitudinal acoustic phason mode and a flat optical phonon mode, both of which cause a significant enhancement in the vibrational density of states. The phasons also acquire a prominent chiral character,  displaying a nonzero angular momentum spread across the Brillouin zone.
We discuss how the tunability of the moir\'e phonon spectra may affect electronic properties, focusing on the recently discovered phenomenon of van der Waals ferroelectricity.
\end{abstract}

\maketitle

\section{Introduction} 
Layered van der Waals (vdW) materials are formed by the assembly of crystalline membranes, combining properties that are characteristic of both \textit{hard} and \textit{soft} condensed matter \cite{Amorim_etal}. 
Graphene, for example, is very stiff within the plane of the honeycomb lattice, but extremely soft with respect to vertical
deformations such as ripples and corrugation, whose 
strain fields can dramatically alter the local electronic properties \cite{Levy2010}.
While encapsulation, epitaxy, or stacking effectively quench these out-of-plane deformations, in these cases a new soft degree of freedom often appears associated with the emergence of incommensurate superstructures: stacking order fluctuations.  

In the case of moir\'e materials, resulting from a lattice mismatch between vdW layers \cite{BM,carr2017twistronics},  the incommensurate moir\'e  superlattice is plagued by soft vibrational modes altering the local stacking arrangement \cite{lin2018moire,Koshino,phasonsI}. 
The role of these moir\'e phonons (and, in particular, of their acoustic modes dubbed \emph{phasons}) on the physics of twisted devices is an area of active research \cite{phasons_TMD,Koshino2020,xiao2021chiral,phasonsII,phasons_Matthias,phasons_Eslam,lu2022low,liu2022moire,phasonsIII,li2023review,Koshino2023}.
Indeed, many moir\'e materials display interesting macroscopic quantum phenomena ranging from incompressible Hall states \cite{nuckolls2020strongly,wu2021chern} to superconductivity \cite{Cao2018a,lu2019superconductors,chen2019signatures,park2021tunable,park2022robust,Xia2024,Guo2024}. One of the reasons why the impact of the moir\'e phonons on these electronic phenomena remains little understood is because of the difficulty in tuning these soft modes by externally controlled parameters. In contrast, the electronic degrees of freedom can be efficiently tuned by electrostatic gating, displacement fields and the twist angle itself \cite{ribeiro2018twistable,inbar2023quantum,tang2023chip}. 

\begin{figure}[t!]
\begin{center}
\includegraphics[width=\columnwidth]{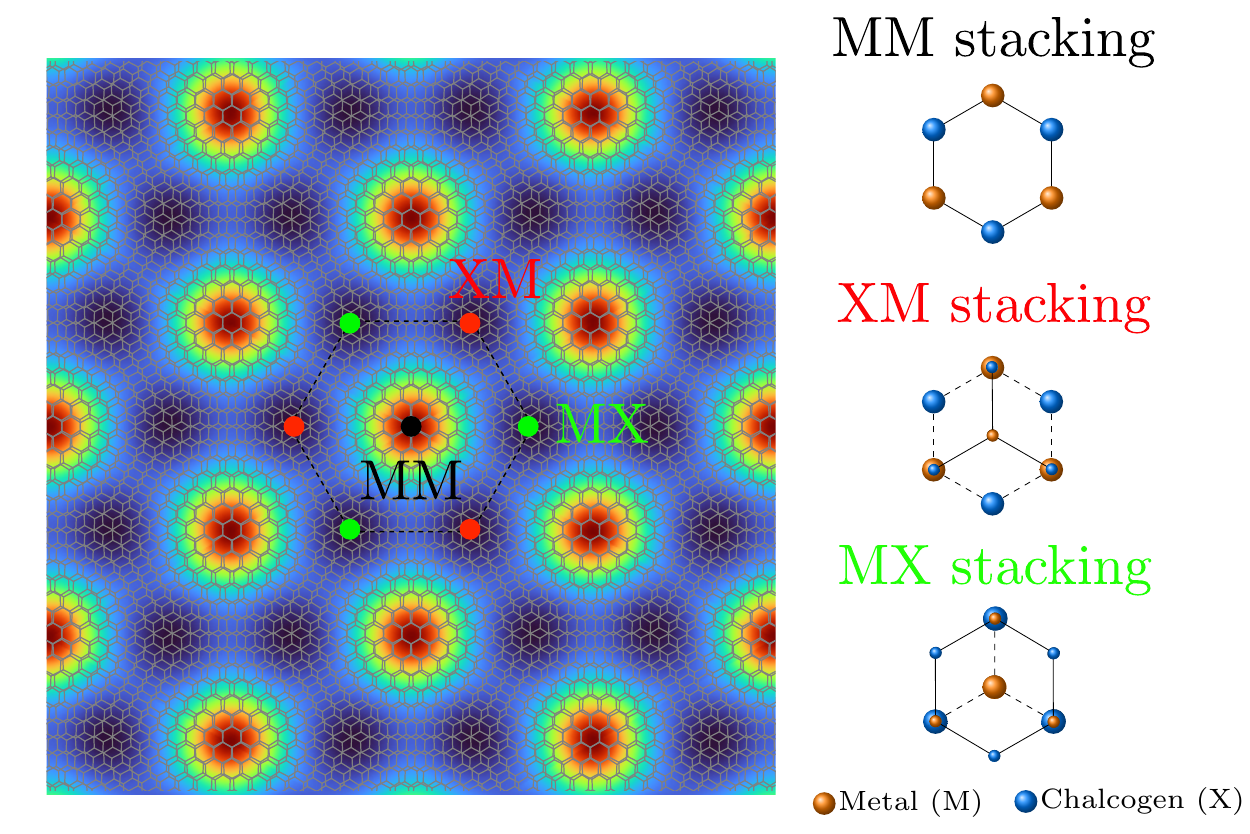}
\caption{\textbf{Stacking configurations in twisted TMD bilayers}. The twist is taken with respect to the non-centrosymmetric MM stacking in which transition-metal (M, orange) and chalcogen (X, blue) sublattices overlap. The resulting moir\'e pattern contains XM and MX stackings connected by mirror reflection (different sizes represent ions on different layers). The local adhesion energy density is superimposed to the twisted lattice. } 
\vspace{-1.cm} 
\label{fig:fig0}
\end{center}
\end{figure}

\begin{figure*}[t!]
\begin{center}
\includegraphics[width=\linewidth]{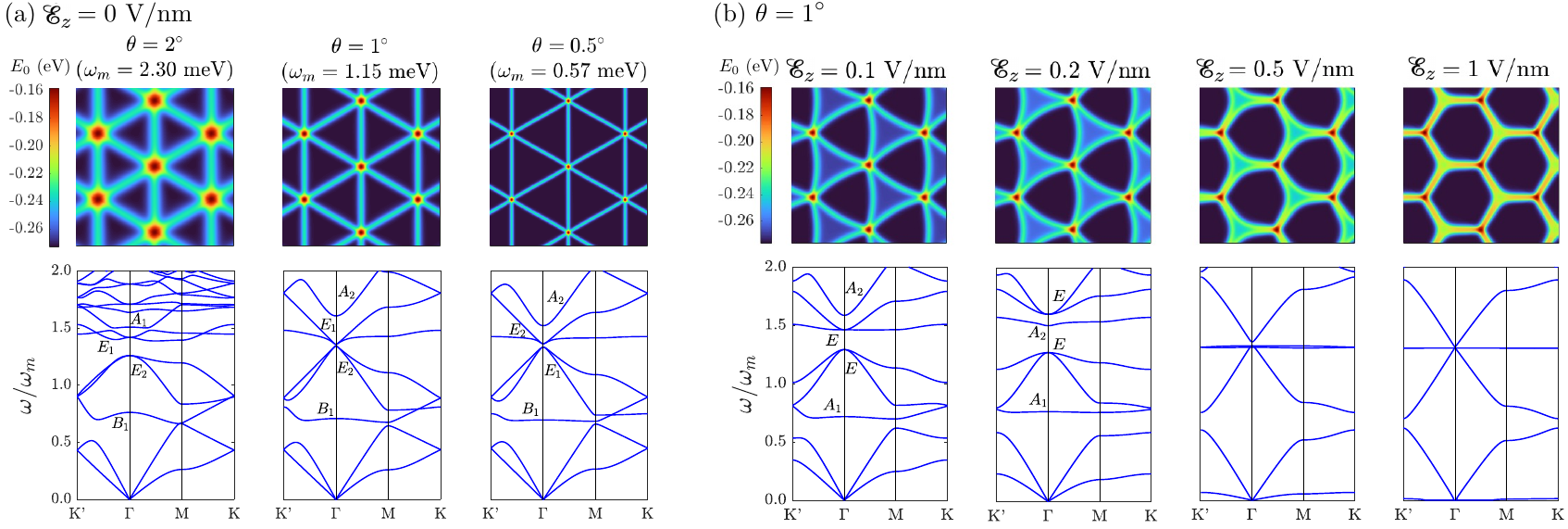}
\caption{\textbf{Stacking relaxation patterns and moir\'e phonon spectra of twisted MoS$_2$}. Top panels depict the adhesion energy density in real space after relaxation. Bottom panels show the phonon band structure in the moir\'e Brillouin zone (mBZ); the scale $\omega_{\textrm{m}}$ is defined in Eq.~\eqref{eq:folding}.  The optical modes at $\Gamma$  are labelled by the irreducible representations of the point group $D_3$ ($D_6$) of the model.
Here and throughout we use $(\mu, \lambda, V_0, V_1, V_2, V_3)$ = (3640, 3330, $-$25.26, 2.76, $-$0.60, 0.18)~meV/\AA$^2$ and $(p_1, p_2, p_3)$ = (0.01375, 0, 0.0030)~e/\AA.}
\vspace{-0.5cm} 
\label{fig:fig1}
\end{center}
\end{figure*}

In addition to correlated phenomena, recent experiments \cite{stern2020interfacial,yasuda2021stacking,wang2022interfacial,weston2022interfacial,ko2023operando,molino2023ferroelectric,van2024engineering,yasuda2024ultrafast,bian2024developing} have revealed ferroelectricity in untwisted and twisted vdW bilayers. These findings have been interpreted in terms of sliding between the layers that changes the global or local stacking order such that inversion symmetry is broken and an out-of-plane polarization emerges  \cite{li2017binary}. In the case of a moir\'e superlattice formed by twisting identical transition metal dichalcogenide  (TMD, formulae MX$_{2}$) layers about a non-centrosymmetric configuration (see Fig.~\ref{fig:fig0}), this mechanism results in an out-of-plane polarization texture defining polar domains tied to the local stacking order \cite{bennett2022electrically,bennett2022theory}. As shown in Fig.~\ref{fig:fig0}, the two types of polar domains are associated with the local stackings MX/XM, in which the metal atom of one layer aligns with the chalcogen atom of the other layer, thus breaking mirror symmetry. The different relative stackings also give rise to in-plane polarization textures with topological character \cite{bennett2023polar,bennett2023theory,vu2024imaging}. Since the polarization couples directly to electric fields, and since the former is tied to the local stacking order, this observation opens a promising path to control the stacking-order fluctuations by an external parameter in moir\'e devices.

In this work, we show how moir\'e phonons in polar moir\'e materials can be tuned using an out-of-plane electric field. Via its coupling to the polar domains, the latter causes commensurate stackings to locally grow and shrink through a bending of the moir\'e domain walls, resulting in a ferroelectric response \cite{yasuda2021stacking,ko2023operando}. Importantly, these changes in the stacking domain configuration dramatically alter the low-energy phonon spectra (see Fig.~\ref{fig:fig1}~b): 
As the symmetry of the moir\'e domain pattern evolves from triangular to honeycomb, the lowest-energy \textit{longitudinal phason} mode of the moir\'e pattern
softens and low-energy optical modes become substantially flattened, enhancing the phonon density of states (DOS). These results provide a pathway to engineer the properties of the moir\'e phonons and thus elucidate their impact on the electronic properties of twisted vdW materials.

\section{Model} We focus on TMD $MX_2$ homobilayers, though
our conclusions can also be extended to heterobilayers \cite{Koshino2}. 
We identify a stacking configuration with a two-dimensional vector $\boldsymbol{\phi}=\boldsymbol{u}_t-\boldsymbol{u}_b$ corresponding to a lateral displacement of the top ($t$) layer with respect to the bottom ($b$) layer. 
The lattice mismatch introduced by twisting causes the stacking configuration to modulate in space. The twist angle is assumed to be small, so that the modulation is smooth on the atomic scale. The corresponding free-energy functional $\mathcal{F}[\boldsymbol{\phi}(\mathbf{r});\mathcal{E}_z]$ depends parametrically on the perpendicular external electric field $\mathcal{E}_z$ via \cite{bennett2022electrically,bennett2022theory}:
\begin{align}
\label{eq:model}
\mathcal{F}\left[\boldsymbol{\phi};\mathcal{E}_z\right]=\mathcal{F}_{\textrm{elas}}\left[\boldsymbol{\phi}\right]+\mathcal{F}_{\textrm{adh}}\left[\boldsymbol{\phi}\right]+\mathcal{F}_{\textrm{elec}}\left[\boldsymbol{\phi};\mathcal{E}_z\right].
\end{align}
The first term represents the usual elastic energy cost associated with heterostrain (repeated indices are summed)
\begin{align}
\mathcal{F}_{\textrm{elas}}\left[\boldsymbol{\phi}(\mathbf{r})\right]=\int d\mathbf{r}\,\left[\frac{\lambda}{4}(\partial_{i}\phi_{i})^{2} + \frac{\mu}{8}(\partial_{k}\phi_{i}+\partial_{i}\phi_{k})^{2}\right] ,
\end{align}
where $\lambda$, $\mu$ are the monolayer Lam\'e coefficients. 
The second term introduces the adhesion energy cost of the different stacking configurations due to the vdW interaction between layers. 
Because by construction $\mathcal{F}\left[\boldsymbol{\phi}+\mathbf{a}\right]=\mathcal{F}\left[\boldsymbol{\phi}\right]$, where $\mathbf{a}$ is a monolayer Bravais vector, the adhesion energy density 
can be expanded in Fourier harmonics,
\begin{align}
{\mathcal{F}}_{\textrm{adh}}\left[\boldsymbol{\phi}\left(\mathbf{r}\right)\right]= \int d\mathbf{r}\, \Big[ V_{0} +\frac{1}{2}\sum_n\sum_{j=1}^{3}\textrm{Re}\left(V_n\, e^{i\boldsymbol{g}_n^{(j)}\cdot\boldsymbol{\phi}\left(\mathbf{r}\right)}\right) \Big].
\end{align}
Hereafter $\boldsymbol{g}_n^{(j)}$, $j=1,2,3$, represent 
monolayer reciprocal lattice vector triplets related by 120$^{\circ}$ rotations. Finally, the third term in Eq.~\eqref{eq:model} represents the coupling of $\mathcal{E}_z$ to the modulating local dipole moments spontaneously formed within the moir\'e supercell \cite{bennett2022electrically,bennett2022theory}, 
\begin{equation}
\begin{split}
\mathcal{F}_{\textrm{elec}}\left[\boldsymbol{\phi}(\mathbf{r});\mathcal{E}_z\right] &= -\int d\mathbf{r}\, p_z\left[\boldsymbol{\phi}(\mathbf{r})\right]\mathcal{E}_z. 
\end{split}
\end{equation}
$p_z\left[\boldsymbol{\phi}(\mathbf{r})\right]$ represents the out-of-plane dipole density $\boldsymbol{p}=p_z\boldsymbol{\hat{z}}$ in stackings breaking mirror reflection and has a similar Fourier expansion: 
\begin{align}
 p_z\left[\boldsymbol{\phi}(\mathbf{r})\right]=\frac{1}{2}\sum_n\sum_{j=1}^{3}\textrm{Im}\left\{p_n\, e^{i\boldsymbol{g}_n^{(j)}\cdot\boldsymbol{\phi}\left(\mathbf{r}\right)}\right\}.
\end{align}

The twist produces a moir\'e pattern containing MM, XM and MX configurations, shown in Fig.~\ref{fig:fig0}. 
MM, which is mirror-symmetric and non-polar, is the reference configuration, $\boldsymbol{\phi}=\vec{0}$. 
The MX/XM stackings are related by mirror reflection, and thus $\mathcal{F}_\mathrm{adh}$ ($p_z$) must be even (odd). Consequently, the coefficients $V_n$ and $p_n$ must be real. They are obtained from first-principles calculations of commensurate bilayers with a relative shift between the layers \cite{bennett2022electrically,bennett2022theory}. $p_n$ are further rescaled to match the experimentally observed field magnitudes (see Supplementary Material~\onlinecite{SM}).

\section{Relaxation patterns}
We first determine the stacking texture $\boldsymbol{\phi}_0(\mathbf{r})$ that minimizes the free energy in Eq.~\eqref{eq:model} subjected to an imposed global twist angle $\theta$ \cite{nam2017lattice,carr2018relaxation,zhang2018structural}; details of the numerical implementations can be found in the Supplementary Material~\onlinecite{SM} (see also Refs.~\onlinecite{NIST:DLMF,nocedal1999numerical,Optim.jl-2018, bezanson2017julia}). Density plots of the adhesion energy landscape of the relaxed structure, $E_0(\mathbf{r})=\mathcal{F}_{\textrm{adh}}\left[\boldsymbol{\phi}_0(\mathbf{r})\right]+\mathcal{F}_{\textrm{elec}}\left[\boldsymbol{\phi}_0(\mathbf{r});\mathcal{E}_z\right]$,  are shown in Fig.~\ref{fig:fig1}  for twisted bilayer MoS$_2$ (see model parameter values in the caption). At zero electric field, Fig.~\ref{fig:fig1}(a), the stacking domain walls (corresponding to solitons of $\boldsymbol{\phi}$) form a triangular network of partial dislocations 
that separates approximately uniform MX and XM configurations (darker areas), intersecting at MM stackings (brighter spots). The degree of relaxation increases as the twist angle decreases, reflecting the ratio between the moir\'e period, $L_{\textrm{m}}\sim a/\theta$, and the characteristic domain-wall width, $\ell \sim a \sqrt{\mu/V_1}$. 

\begin{figure}[t!]
\begin{center}
\includegraphics[width=\columnwidth]{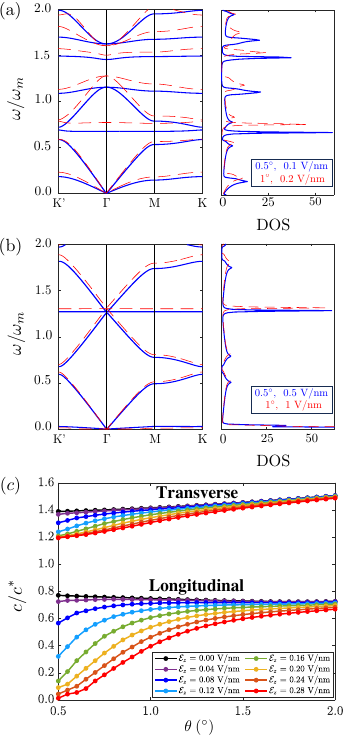}
\caption{\textbf{Universal features of the moir\'e phonon spectrum}. (a) Phonon bands and density of states (DOS) per cell (in units of $\omega_{\textrm{m}}^{-1}$) for two different angles and fields (but same ratio $L_{\textrm{m}}/R$). (b) The same as in (a) but for larger fields corresponding to a honeycomb soliton network. (c) Evolution of the transverse and longitudinal phason velocities with twist angle for different values of $\mathcal{E}_z$. In the vertical axis, $c^{*} = \sqrt{\mu/\rho}$ is the transverse sound velocity of the monolayer.  As the field increases, the curves approach each other for the smallest angles.} 
\vspace{-1.cm} 
\label{fig:fig3}
\end{center}
\end{figure}

In the presence of an electric field, Fig.~\ref{fig:fig1}~(b), the MX and XM domains become energetically inequivalent. Consequently, one domain expands and the other contracts, bending the solitons. For small angles, the domain walls can be envisioned as strings under a linear tension $\sigma$ imposed by the adhesion potential with curvature radius $R$ given by \cite{SM} 
\begin{align}
\label{eq:kappa}
R \sim\frac{\sigma}{3\sqrt{3}|\mathcal{E}_z|(p_1-p_3)}.
\end{align}
As the field increases, $R$ approaches the moir\'e period $L_{\textrm{m}}$ and the solitons merge into perfect screw dislocations \cite{enaldiev2022scalable}. 
For strong electric fields, $R \ll L_{\textrm{m}}$, the relaxation pattern evolves into a honeycomb soliton network (see last panel in Fig.~\ref{fig:fig1}~b), fundamentally changing the stacking fluctuation spectrum. 

\section{Moir\'e phonons}
Diagonalization of the dynamical matrix defined by the harmonic expansion of $\mathcal{F}$ in the stacking fluctuations $\delta\boldsymbol{\phi}(\mathbf{r},t)=\boldsymbol{\phi}(\mathbf{r},t)-\boldsymbol{\phi}_0(\mathbf{r})$ gives the moir\'e phonon spectra. For twisted MoS$_2$, they are shown in Fig.~\ref{fig:fig1}  together with the corresponding relaxation patterns. The characteristic frequency
\begin{align}
\label{eq:folding}
\omega_{\textrm{m}}=\sqrt{\frac{\mu}{3\rho}}\frac{4\pi}{L_{\textrm{m}}},
\end{align}
where $\rho=3\cdot10^{-6}$ kg/m$^2$ is the monolayer mass density, is of the order of 1 meV (THz range) around $\theta=1^{\circ}$. 

Figure~\ref{fig:fig1}~(a) displays the phonon spectra in the moir\'e Brillouin zone (mBZ) for different twist angles at zero field. The two linearly dispersing acoustic-like branches emanating from the $\Gamma$ point correspond to the sliding phasons of the relaxation pattern. Contrary to a crystalline lattice, the acoustic mode with the lowest energy is the longitudinal one \cite{phasonsI}. Both phason velocities are approximately independent of the twist angle \cite{Koshino,phasonsI,phasons_Eslam}, and closely match the corresponding monolayer speeds of sound (see black curve in Fig.~\ref{fig:fig3}~c).

The folding of the monolayer acoustic phonons into the mBZ gives rise to new optical phonon modes.
The sharpening of the domain walls at small angles is manifested in the gradual flattening of some of the optical branches, see Fig.~\ref{fig:fig1}(a). The optical modes at $\Gamma$ are labeled according to the irreducible representations of the $D_6$ point group of the model ($D_3$ for $\mathcal{E}_z \neq 0$), and can be understood as different standing-wave patterns of the domain walls \cite{Koshino,Koshino2023}.  
Fig.~\ref{fig:fig1}~(b) shows the evolution of the phonon spectrum with increasing $\mathcal{E}_z$ for fixed $\theta=1^{\circ}$. Moderate fields introduce appreciable changes, such as the inversion between singlet ($A_2$) and doublet ($E$) modes at the $\Gamma$ point for $\mathcal{E}_z$ between 0.1 and 0.2 V/nm.

As the field increases further, the spectrum approaches a universal (i.e. $\theta$-independent) shape that is well-described by the model of strings under tension of Ref.~\onlinecite{Koshino2023}. Figures~\ref{fig:fig3}(a)~and~(b) further illustrate this point by comparing the spectra of two different angles and fields with the same value of $L_{\textrm{m}}/R$. In the honeycomb regime [panel (b)], the two phonon spectra nearly overlap and display very flat optical branches, which are manifested as sharp peaks in the vibrational DOS. Note also that in the limit $\mathcal{E}_z \to \infty$ the honeycomb soliton network recovers an emergent $C_6$ symmetry, leading to the eventual closing of some of the gaps at the K points (e.g. around $\omega/\omega_m\approx0.6$ in Fig.~\ref{fig:fig3}b). 

\section{Phason softening}
The effect of the electric field on the acoustic phason modes, and the lowest-energy longitudinal mode in particular, is striking. As shown in Figs.~\ref{fig:fig1}(b) and ~\ref{fig:fig3}, the longitudinal phason band becomes almost completely flat in the limit of large $\mathcal{E}_z$ , leading to a sharp DOS peak close to $\omega=0$. Indeed, Fig.~\ref{fig:fig3}~(c) demonstrates that the longitudinal velocity is strongly softened by the electric field, deviating substantially from the monolayer value. This effect can be partly attributed to the $\mathcal{E}_z$-induced opening of gaps in the phonon spectrum at the mBZ corners, which causes level repulsion between phonon branches.

More broadly, the existence of such a nearly zero-velocity acoustic mode in a honeycomb soliton network is not unexpected. In a system of sharp domain walls, the energy is essentially controlled by the wall length. Interestingly, in the honeycomb case, displacements of network vertices that preserve the wall orientations and connectivity leave the total length of the soliton network invariant \cite{Villain}. 
This makes it possible to change the area of the hexagons with little energy cost, leading to a soft longitudinal phason.

\begin{figure}[t!]
\begin{center}
    \includegraphics[width=\columnwidth]{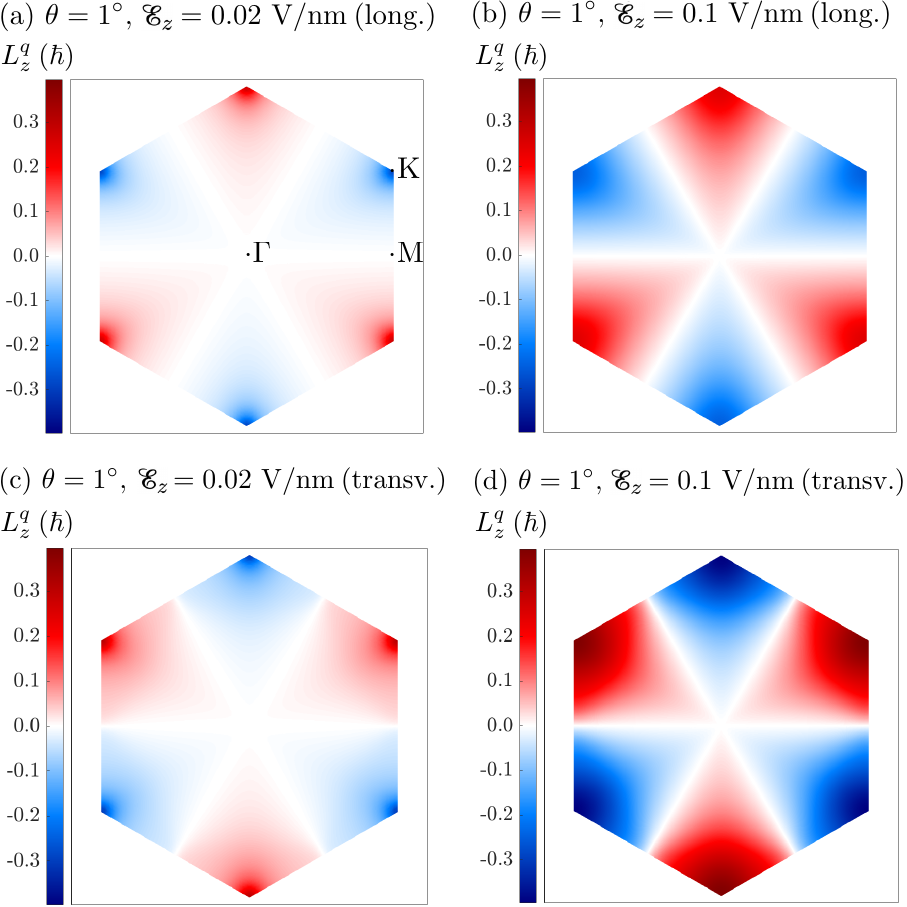}
    \caption{\textbf{Angular momentum of the acoustic phasons of MoS$_{2}$ within the mBZ for varying electric field}. The plots show the value of the angular momentum $L_{z}^{\boldsymbol{q}}$ for a single twist angle $\theta = 1^{\circ}$. The first and second row correspond to the longitudinal  and transverse acoustic modes, respectively. Each column corresponds to a different value of $\mathcal{E}_{z}$.} 
    \vspace{-0.5cm} 
\label{fig:fig4}
\end{center}
\end{figure} 

\section{Phonon angular momentum}
Because the electric field breaks the $C_{2z}$ symmetry of the model, moir\'e phonon bands may acquire nonzero angular momentum 
$ L^{\boldsymbol{q}}_{z}= i\hbar\sum_{\boldsymbol{G}}\delta\boldsymbol{\phi}_{\boldsymbol{q}+\boldsymbol{G}} \times \delta\boldsymbol{\phi}_{\boldsymbol{q}+\boldsymbol{G}}^{*}$
perpendicular to the bilayer \cite{Koshino2023, xiao2021chiral}; here $\boldsymbol{\phi}_{\boldsymbol{q}+\boldsymbol{G}}$ are the Fourier components of the fluctuation mode (the band index is omitted) \cite{SM}. Although these are often called \textit{chiral phonons} \cite{QianNiu2015}, note that modes at opposite points of the mBZ carry opposite angular momentum due to time-reversal symmetry. Figure~\ref{fig:fig4} displays the distribution of $L^{\boldsymbol{q}}_{z}$ for the acoustic modes within the mBZ for fixed $\theta = 1^{\circ}$ and two values of $\mathcal{E}_z$. While for small fields the angular momentum is concentrated around the K points, it spreads across the mBZ as $\mathcal{E}_{z}$ increases. Moreover, its magnitude depends on the energy of the phason modes at the K points. 

Phason chirality, understood as finite but opposite values of $L_{z}^{\boldsymbol{q}}$ at inequivalent K points, can be traced back to complex conjugate eigenvalues of phason modes under $C_{3}$ rotations at those points (sometimes called pseudo-angular momentum \cite{QianNiu2015}), which gives rise to different selection rules with circularly polarized light.
In untwisted 2D materials, chiral phonons have been observed in transient infrared spectroscopy and Raman experiments \cite{chen2018chiral, chen2015helicity, zhang2018chiral, zhang2023weyl}. For twisted moir\'e systems, the chirality and the softening of the acoustic phasons could in principle be measured using Brillouin-Mandelstam spectroscopy. In this technique, the incidence angle of light can be changed to probe phonons with different wavevectors, thus providing a measurement of the velocity of the modes \cite{Kargar2021}. 

\section{Discussion}
Here, we showed that the low-energy moir\'e phonons of twisted bilayers composed of materials lacking a center of inversion are strongly influenced by the application of an out-of-plane electric field. 
The electric field couples to the polarization density which in turn is tied to the stacking texture, therefore changing the adhesion energy landscape and subsequently the geometry of the stacking relaxation patterns. 
These are described by three characteristic length scales: the moir\'e period $L_{\textrm{m}}$ imposed by the twist, the domain wall width $\ell$ generated by the adhesion potential, and the curvature radius $R$ [Eq.~\eqref{eq:kappa}] of the walls introduced by the electric field. 

While the strongest effect of the field is found in the phason (acoustic) modes describing traveling waves of the domain wall system, there are noticeable effects in some of the low-energy optical modes. 
Gap openings at the mBZ corners due to the reduction of the symmetry contribute to better resolve them in energy, giving rise to sharper features in the vibrational DOS. 
The electric field also alters the energetic ordering of optical modes with different symmetries, such as the aforementioned crossing of singlet and doublet modes at the $\Gamma$ point in Fig.~\ref{fig:fig1}~(b). 
Some of these phonon energy shifts could be observed in optics experiments, since the doublet is both infrared and Raman active due to the lack of an inversion center in the structure.

The moir\'e phonons acquire a universal ($\theta$-independent) spectrum in the limit of large fields, $ L_{\textrm{m}}\gg R$, when the geometry of the relaxation pattern changes from a triangular to a honeycomb soliton network. This behavior is attributed to the fact that the phonon energetics are completely dominated by the length of the domain walls. In this regime,
several phonon modes become flat. 
The flat optical phonons can be understood as standing waves of the domain walls subjected to the new boundary conditions of the field-deformed domain-wall network \cite{Koshino2023}. 
On the other hand, the flattening of the longitudinal phason mode is a characteristic feature of the properties of the honeycomb soliton network. Importantly, reaching this regime of large fields is experimentally feasible ($\mathcal{E}_z=0.5$ V/nm for $\theta=0.5^{\circ}$, see Fig.~\ref{fig:fig3}~b). 

In what concerns the interplay with moir\'e ferroelectricity, phonons are believed to strongly renormalize the ferroelectric switching field and depolarization temperature in aligned bilayer \cite{tang2023sliding}. It will be interesting to elucidate if/how the phonon softening impacts this behavior. More broadly, the softening of the longitudinal phasons suggests that electric fields contribute to making the moir\'e pattern with a well-defined twist angle more unstable. 
Nevertheless, at non-zero temperatures and large electric fields, the moir\'e pattern can still be stable due to the large configurational entropy of the honeycomb soliton network \cite{Villain}. 
In this regard, the phonon spectra of entropically stablized moir\'e patterns should acquire a strong dependence on temperature that is not captured by our theory. 

In conclusion, in polar moir\'e materials, electric fields can be used to tune the flatness of the phonon dispersion, akin to how the twist angle and the displacement field are used to flatten the electronic dispersion. Because the flat optical and phason modes give rise to sharp peaks in the low-energy vibrational DOS, they should be manifested in the specific heat and thermal conductivity. Moreover, the electric field should also affect electronic properties that are directly impacted by the electron-phonon coupling. In this regard, the prospect of experimentally controlling the moir\'e phonon spectrum provides a unique opportunity to elucidate whether phonons play a prominent role in promoting some of the electronic phenomena observed in twisted moi\'e systems, such as superconductivity \cite{Xia2024,Guo2024} or linear-in-$T$ resistivity  \cite{phasonsIII}.


\begin{acknowledgments}
\section{Acknowledgements} 
R.M.F. was supported by the U.S. Department of Energy, Office of Science, Basic Energy Sciences, Materials Science and Engineering Division, under Award No. DE-SC0020045.
D.B.~acknowledges the US Army Research Office (ARO) MURI project under grant No.~W911NF-21-0147 and the Simons Foundation award No.~896626. 
B.R.~ was supported by the Yale Postdoctoral Prize fellowship.
\end{acknowledgments}

\bibliography{references}


\clearpage
\newpage

\setcounter{secnumdepth}{5}

\renewcommand{\thefigure}{S\arabic{figure}}
\setcounter{figure}{0} 

\appendix

\onecolumngrid

\begin{center}
{\large \bf Supplemental Information: Flat and tunable moir\'e phonons in twisted transition-metal dichalcogenides} 
\end{center}

\addtocontents{toc}{\protect\setcounter{tocdepth}{-1}}


\subsection{Estimation of the model parameters}

Initial values for $V_n$ and $p_n$ parametrizing ${\mathcal{F}}_{\textrm{adh}}\left[\boldsymbol{\phi}\left(\mathbf{r}\right)\right]$ and $p_z\left[\boldsymbol{\phi}(\mathbf{r})\right]$ in the main text were taken from first-principles calculations in Ref.~\cite{bennett2022theory}. From these numbers we obtain a coercive field for polarization switching in commensurate stackings of $\mathcal{E}_{\rm c} = 13.3$ V/nm.
It has been put forward that the associated bare energy barrier height is strongly renormalized downwards by vibrational fluctuations \cite{tang2023sliding}, which this estimate does not include. In addition to that, first-principles calculations tend to overestimate the coercive electric field values in ferroelectric materials. The typical observed coercive fields in transition metal dichalcogenides are approximately 0.15 V/nm \cite{ko2023operando}.

We therefore apply a scaling correction to the dipole coefficients, $p_n \to \chi p_n$, in order to align our results with experiment. 
A simple estimate of the coercive field is the electric field required to overcome the energy barrier. It can be obtained as follows: From the energy of a commensurate, rigid displacement,
\begin{align}
 F_0(\vec{\phi}; \mathcal{E}_z) = {\mathcal{F}}_{\textrm{adh}}\left[\vec{\phi}\right] + {\mathcal{F}}_{\textrm{elec}}\left[\vec{\phi }; \mathcal{E}_z \right],
\end{align}
we determine the field value $\mathcal{E}_z$ at which
\begin{align}
 E_0(\vec{\phi}_{\rm MX};\mathcal{E}_z) = E_0(\vec{\phi}_{\rm DW}; \mathcal{E}_z),
\end{align}
where $\vec{\phi}_{\rm MX} = (\vec{a}_1 + \vec{a}_2)/3$ (one third of a unit cell diagonal) corresponds to the MX stacking, and $\vec{\phi}_{\rm DW} = (\vec{a}_1 + \vec{a}_2)/2$ (one half of a unit cell diagonal) corresponds to the high-energy intermediate stacking in a domain wall; here $\vec{a}_{1,2}$ are primitive vectors of the Bravais lattice.

In this way, a coercive field of $\mathcal{E}_{\rm c} = 0.15$ V/nm is achieved by taking $\chi = 89$. The rescaled parameters we use are provided in the caption of Fig.~\ref{fig:fig1} in the main text. The qualitative physics in this work are not affected by this correction, but this allows us to make better comparisons with experimental measurements, in particular with the relaxation patterns as a function of electric field~\cite{ko2023operando}.

\subsection{Symmetries of the model}

The spatial symmetries of a twisted TMD bilayer are contained in a chiral point group symmetry no larger than $D_3$. However, for $\mathcal{E}_z=0$ our model has a larger 6-fold symmetry owing to the fact that the adhesion energies of stacking configurations related by mirror reflection (e.g., XM and MX stackings) are the same. The 3-fold rotational symmetry is restored by $\mathcal{E}_z\neq 0$, however, the model still retains spurious C$_2$-rotation axes (or, for a planar group, mirror reflection planes) within the plane of the sample, despite the fact that the actual symmetry is the chiral and polar group $C_3$. 

For this reason, in the main text  we label the phonon modes by irreducible representations (irreps) of $D_6$ in the absence of electric fields (or $C_{6v}$, despite that the actual symmetry of the system is $D_3$) and in the presence of an electric field, as irresps of $D_3$ (or $C_{3v}$, despite that the actual symmetry of the system is $C_3$ in that case). 

The spurious symmetries introduced in the model are ultimate consequence of our coarse-graining in which the two layers are not treated individually: By construction, the displacements on each layer are always of the same magnitude but in opposite directions.

\subsection{Moir\'e relaxation}

\begin{figure}[t!]
\begin{center}
    \includegraphics[width=0.5\linewidth]{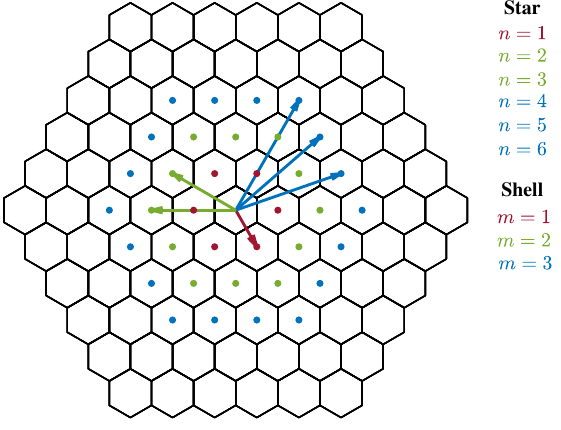}
    \caption{\textbf{Harmonics in reciprocal space}. The central hexagon represents the first moir\'e Brillouin zone (mBZ), and the honeycomb lattice is obtained translating the mBZ by moir\'e reciprocal lattice vectors $\vec{G}$. Each color represents a shell, whereas each vector spans a star by consecutive $C_{6}$ rotations. Note that the $m-$th shell contains $m$ stars and that each star consists of six harmonics. In the absence of an electric field all the harmonics within a star are related by $C_{6}$ symmetry. For nonzero fields only $C_{3}$ symmetry is preserved and therefore each star contains two groups of three harmonics related by $C_{3}$ rotations.} 
\vspace{-0.5cm} 
\label{fig:S0}
\end{center}
\end{figure}

In the relaxation problem we look for the stacking field $\vec{\phi}_{0}(\vec{r})$ that minimizes the free energy of the bilayers rotated by a twist angle $\theta$. We decompose $\vec{\phi}_0$ as
\begin{align}
\boldsymbol{\phi}_0(\vec{r})=2\sin\frac{\theta}{2}\,\boldsymbol{\hat{z}}\times\vec{r}+\boldsymbol{u}_0(\vec{r}),
\end{align}
where the first term encodes a rigid twist and the second represents relaxation.
The relaxation displacement field $\boldsymbol{u}_0(\mathbf{r})$ is periodic with the moir\'e periodicity, and is spanned by
\begin{align} \label{eq:u_fourier}
    \vec{u}_{0}(\vec{r}) = \sum_{\vec{G}}\vec{u}_{\vec{G}}\: e^{i\vec{G}\cdot\vec{r}},
    \end{align}
where $\vec{G}=2\sin\frac{\theta}{2}\,\vec{g}\times\boldsymbol{\hat{z}}$ are the moir\'e reciprocal lattice vectors and $\vec{g}$ are the 
monolayer reciprocal lattice vectors (see Fig. \ref{fig:S0}).

We determine $\boldsymbol{u}_0(\mathbf{r})$ by two procedures. In the first method we solve the saddle-point equations by iteration. In the second procedure, suitable for smaller angles and larger fields, we treat the Fourier coefficients $\vec{u}_{\vec{G}}$ as variational parameters and minimize $\mathcal{F}$ directly by gradient descent \cite{bennett2022electrically,bennett2022theory}. 

\subsection{Iterating saddle-point equations}

The Euler--Lagrange equations derived from Eq.~\eqref{eq:model} are:
\begin{align}
     \frac{\lambda+\mu}{2}\vec{\nabla}(\vec{\nabla}\cdot\vec{u}_{0}) + \frac{\mu}{2}\vec{\nabla}^{2}\vec{u}_{0} = \left.\frac{\partial \bar{\mathcal{F}}_{\textrm{adh}}}{\partial\vec{\phi}} \right|_{\vec{u}=\vec{u}_{0}} + \left.\frac{\partial \bar{\mathcal{F}}_{\textrm{elec}}}{\partial\vec{\phi}}\right|_{\vec{u}=\vec{u}_{0}}.
\end{align}
It is convenient to write $\vec{u}_{\vec{G}}$ in longitudinal and transverse components,
\begin{align}
    \vec{u}_{\vec{G}} = \frac{\vec{G}}{|\vec{G}|}\,u_{\vec{G}}^{L} + \frac{\vec{G}\times\vec{\hat{z}}}{|\vec{G}|}\,u_{\vec{G}}^{T}.
\end{align}
In this notation the equations read
\begin{align} \label{eq:numerics}
&    \frac{3}{4\pi d}\,u_{\vec{G}}^{L} = 
    \frac{\mu}{2\mu+\lambda}\frac{1}{L_{m}^{2}|\vec{G}|^{2}}
    \sum_{n,j} \left(\frac{\vec{G}}{|\vec{G}|} \cdot \frac{\vec{g}_{n}^{(j)}}{|\vec{g}_{1}^{(1)}|}\right)F_n^{(j)},\qquad
    &\frac{3}{4\pi d}\,u_{\vec{G}}^{T} = 
    \frac{1}{L_{m}^{2}|\vec{G}|^{2}}
    \sum_{n,j} \left(\frac{\vec{g}_{n}^{(j)}}{|\vec{g}_{1}^{(1)}|} \times \frac{\vec{G}}{|\vec{G}|}\right)_zF_n^{(j)},
\end{align}
where $n$ runs over momentum stars and $j$ over elements of the star (see Fig.~\ref{fig:S0} for notation), and we have introduced dimensionless forces
\begin{align} \label{eq:forces}
F_n^{(j)}=\frac{1}{A_{m}} 
    \int d\vec{r}\, \bigg[ 
    \left(\frac{L_{m}}{\ell_{n}^{\textrm{even}}}\right)^{2}
    \sin(\vec{G}_{n}^{(j)}\vec{r} + \vec{g}_{n}^{(j)}\vec{u}_{0}) 
    -  \left(\frac{L_{m}}{\ell_{n}^{\textrm{odd}}}\right)^{2}
    \cos(\vec{G}_{n}^{(j)}\vec{r} + \vec{g}_{n}^{(j)}\vec{u}_{0})
    \bigg] 
    e^{-i\vec{G}\vec{r}} \: ,
\end{align}
where the integral is on a moir\'e cell with area $A_{m}$. Finally, we have introduced the characteristic lengths
\begin{align}
    & \ell_{n}^{\textrm{even}}=d\sqrt{\frac{\mu}{2V_{n}}}, 
    \qquad  \ell_{n}^{\textrm{odd}}=d\sqrt{\frac{\mu}{2p_{n}\mathcal{E}_{z}}},
\end{align}
where $d=1.8267$ $\AA$ is the interatomic distance in one layer. 

We seed our iterative scheme with $\vec{u}_{\vec{G}} = \vec{0}$ for all $\vec{G}$ and $F_n^{(j)} = 0$ for all $n,j$. Equations~\eqref{eq:forces} and \eqref{eq:numerics} are then iterated to compute new values for $\vec{u}_{\vec{G}}$ and $F_n^{(j)}$. The process continues until a fixed point is reached. 

\subsection{Optimization by gradient descent}
Capturing the structure of the domains walls requires retaining Fourier coefficients up to wavevectors $|\vec{G}| \sim  1 / \ell$, where $\ell$ is the characteristic domain-wall width (see below). The number of momentum shells required is then $\sim L_m / \ell$, and hence the number of momentum stars involved is $(L_m \ell)^2 \sim (V_1 + p_1 \mathcal{E}_z) / \mu \theta^2$. Hence, in the limit of small angles and large electric fields, the number of necessary Fourier coefficients grows rapidly. The foregoing iterative method is ill-suited for such a large number of variables. 

Instead, we compute the relaxation pattern by numerical optimization of the free energy,  Eq.~\eqref{eq:model} of the main text. 
Truncating at $N$ momentum shells, there are $M = N (N+1) / 2$~~$C_6$-unique momentum stars. We collect the $2M$ components $\vec{u}_{\vec{G}}$  in a length-$2M$ vector $\mathsf{U}$. The linearity of the Fourier decomposition \eqref{eq:u_fourier}
implies that at fixed $\vec{r}$, $\vec{u}_0 (\vec{r}) = \mathsf{C}(\vec{r}) \cdot \mathsf{U}$, where $\mathsf{C}(\vec{r})$ is a $2\times 2M$ matrix of position-dependent coefficients. The free energy can then be expressed as
\begin{equation} \label{eq:energy-vectorized}
\mathcal{F}[\mathsf{U}] = \sum_{\vec{G}}    \left[\frac{\mu}{4} G^2 |\vec{u}_{\vec{G}}|^2 + \frac{\mu + \lambda}{4}|\vec{G} \cdot \vec{u}_{\vec{G}}|^2 \right] 
+ \int\mathcal{F}_{\mathrm{adh+elec}}(\mathsf{C}(\vec{r}) \cdot \mathsf{U})  d^2\vec{r} 
= \frac{1}{2} \mathsf{U} \cdot \mathsf{F}_{\mathrm{elas}} \cdot \mathsf{U} +  \sum_{i} \mathcal{F}_{\mathrm{adh+elec}}(\mathsf{C}(\vec{r}_i) \cdot \mathsf{U}) w_{i}.
\end{equation} 
Here the first term gives the elastic energy contribution (in Fourier space), expressed as a bilinear in $\mathsf{U}$. The second term is the nonlinear adhesion energy integral, which runs over one moir\'e unit cell. In the last expression the integral is discretized to a grid of positions $\vec{r}_i$ and integration weights $w_i$ by Gauss--Legendre quadrature \cite{NIST:DLMF}. 

Equation \eqref{eq:energy-vectorized} makes explicit the free energy gradient with respect to the variational parameters,
\begin{equation}
    \nabla_{\mathsf{U}} \mathcal{F} = \mathsf{F}_{\mathrm{elas}} \cdot \mathsf{U} +  \sum_{i} \nabla_{\vec{u}_0}\mathcal{F}_{\mathrm{adh+elec}}(\mathsf{C}(\vec{r}_i) \cdot \mathsf{U}) \cdot \mathsf{C}(\vec{r}) w_i,
\end{equation}
making the problem well-suited to gradient-descent methods. 
We optimize the energy functional  using the Limited-memory Broyden--Fletcher--Goldfarb--Shanno (L-BFGS) algorithm \cite{nocedal1999numerical}, implemented in the \texttt{Optim.jl} package of the \texttt{Julia} programming language \cite{Optim.jl-2018, bezanson2017julia}.  

We begin the descent from the strain-free configuration $\mathsf{U} = \vec{0}$. The relaxation solution is not unique (evidenced by the gapless acoustic phason modes), and so to pick a unique solution we fix the coefficients corresponding to $\vec{u}_{\vec{G}=\vec{0}} = \vec{0}$ identically. We incrementally obtain the relaxation solution $\mathsf{U}$ for a smaller number $N'<N$ of shells, which is then the starting point for a new gradient descent search in the enlarged phase space corresponding to $N'+1$ shells. This is repeated until we reach the final desired shell number $N$. At each iteration, the convergence criterion is specified in terms of the gradient magnitude $|\nabla_{\mathsf{U}} \mathcal{F}|$.

\subsection{Numerical implementation}

We have included 10 momentum shells (55 stars) to produce the relaxation and phonon dispersion figures for $\theta=1^{\circ}$ and $\mathcal{E}_{z}\leq 0.2$ V/nm, and 17 shells (153 stars) for larger fields or $\theta = 0.5^{\circ}$. 

To determine the velocity $c$ of the acoustic modes we take the arithmetic mean of the velocities along $\Gamma-$M and $\Gamma-$K. The reason is that for each phason $c$ should be isotropic around $\Gamma$, in a region whose size we do not control, so in this way we get a better estimation of the actual value. For each one, for instance $c_{\Gamma-\text{M}}$ corresponding to the path $\Gamma-$M, we calculate $c=\Delta\omega/\Delta q$, with $\Delta \omega = \omega(\text{M}/\xi)-\omega(\Gamma) = \omega(\text{M}/\xi)$ and $\Delta q = |M|/\xi$. The factor $\xi$ is a numerical parameter that we tune for each electric field, as the linearity of the dispersion within the mBZ survives until different distances away from $\Gamma$ for different values of $\mathcal{E}_{z}$. In order to remove noise from the values of the velocities shown in Fig.~\ref{fig:fig3} we used a greater amount of shells (32 for the smallest angles and greatest fields) and the gradient descent method.

\begin{figure}[t!]
\begin{center}
    \includegraphics[width=0.5\linewidth]{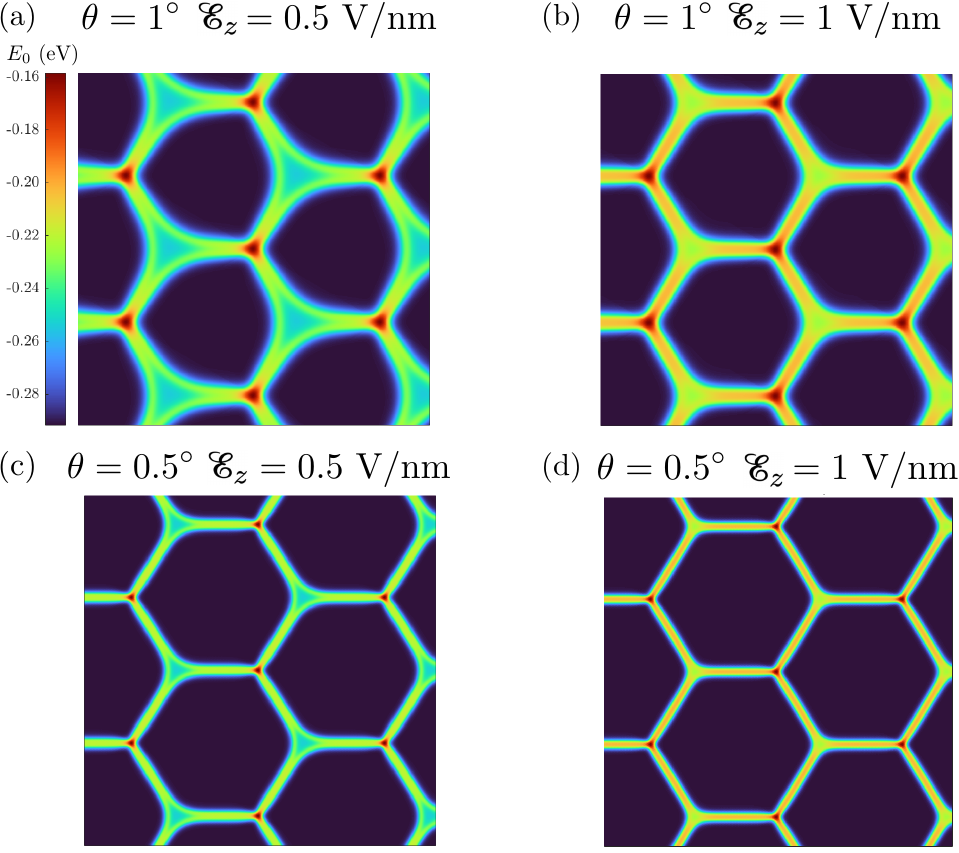}
    \caption{\textbf{Relaxation patterns with large electric fields}. Panels (a) and (b) show the potential landscape at $\theta=1^{\circ}$ and two values of the electric field. Similar for panels (b) and (c), for $\theta = 0.5^{\circ}$. Weaker electric fields are needed to enter the honeycomb regime  for smaller twist angles. In all four cases we have used 17 momentum shells (153 symmetry-distinct $\vec{G}$ vectors) in the numerical calculations.} 
\vspace{-0.5cm} 
\label{fig:S1}
\end{center}
\end{figure}

\subsection{Length scales of the relaxation patterns}

In the limit of small twist angles, $L_{m}\gg \ell_n^{\textrm{even}},\ell_n^{\textrm{odd}}$, the relaxed moir\'e pattern can be envisioned as a network of solitons/domain walls/strings under tension. In the absence of a field, $\mathcal{E}_z=0$, the solitons are well described by pure-shear sine-Gordon domain walls of characteristic width\begin{align}
\ell=\frac{d}{2\pi}\sqrt{\frac{\mu}{V_1-8V_2+9V_3}},
\end{align}
which connect energetically equivalent regions of partial commensuration, XM and MX stackings, and intersect at MM configurations forming the triangular network in panel~(a) of Fig.~2 in the main text. Note that $\ell$ is independent of the twist angle, thus for smaller angles the domain walls become sharper on the moir\'e scale. In this limit the energy of the relaxation pattern is governed by the length of the domain walls, which can be thought as a system of strings under linear tension\begin{align}
\sigma=2\ell (V_1-8 V_2+9V_3).
\end{align}

As shown in Fig.~2~(b) of the main text, for small values of the field $\mathcal{E}_z$ the domain walls are bent as result of the forces produced by the energy imbalance between MX and XM stackings. If a domain wall is extended along direction $x$, we can describe its deformation by a displacement field $u(x)$, so that a point corresponding to the saddle point between the two minima is located at position $(x,u(x))$ in the plane. The curvature $\kappa$ of the domain wall is defined as
\begin{align}
    \kappa= \frac{u''}{[1 + (u')^{2}]^{3/2}}.
\end{align}
At small fields the shape of the domain wall can be described as the one arising from a force on an isolated string of total length $L_{\textrm{m}}$ subjected to fixed boundary conditions, $u(\pm L_{\textrm{m}}/2)=0$. The energy functional of the string reads
\begin{align}
    E[u(x)] = \sigma \int dx\,\sqrt{1+(u')^2}\: - (E_{\textrm{MX}}-E_{\textrm{XM}})\int dx\, u(x).
\end{align}
The first term represents the energy cost of changes in the length of the domain wall. The second term introduces the force due to the energy imbalance between MX and XM stackings; in our model:\begin{align}
E_{\textrm{MX}}-E_{\textrm{XM}}=3\sqrt{3}\mathcal{E}_z\left( p_1-p_3\right).
\end{align}
Minimizing this functional, we arrive at\begin{align}
E_{\textrm{MX}}-E_{\textrm{XM}}+\sigma\frac{u''}{[1 + (u')^{2}]^{3/2}}=0\Rightarrow \kappa=-\frac{3\sqrt{3}\mathcal{E}_z\left(p_1-p_3\right)}{2\ell (V_1-8V_2+9V_3)}.
\end{align}

The inverse of $|\kappa|$ defines the curvature radius $R$ of the domain wall introduced in the main text. For $|\kappa|L_{\textrm{m}}\ll 1$ the deformation is small, domain walls do not overlap and their shape is well approximated by the solution of the previous equation subjected to fixed boundary conditions:\begin{align}
u(x)=\frac{2\sqrt{1-\kappa^2x^2}-\sqrt{4-\kappa^2L_{\textrm{m}}^2}}{2\kappa}.
\end{align}
The relative change in area of MX and XM stackings as a function of field $\mathcal{E}_z$ can be estimated as\begin{align} 
\Delta A=6\,\int_{-\frac{L_{\textrm{m}}}{2}}^{\frac{L_{\textrm{m}}}{2}} dx \,u(x)=6\,\kappa^{-2}\arcsin\left(\frac{\kappa L_{\textrm{m}}}{2}\right)-\frac{3L_{\textrm{m}}}{\kappa}\sqrt{1-\frac{\kappa^2L_{\textrm{m}}^2}{4}}\approx \frac{\kappa L_{\textrm{m}}^3}{2}.
\end{align}
This expression describes well the measurements in Ref.~\onlinecite{ko2023operando} at small fields (with no hysteresis), which offers another justification for our rescaling of $p_n$ parameters.

In the opposite regime, $|\kappa|L_{\textrm{m}}\gg1$, domain walls overlap and eventually form a honeycomb-like network (Fig.~\ref{fig:S1}) connecting energetically disfavored stacking configurations.


\subsection{Dynamical matrix and phonon bands}
\begin{figure}[t!]
\begin{center}
    \includegraphics[width=\columnwidth]{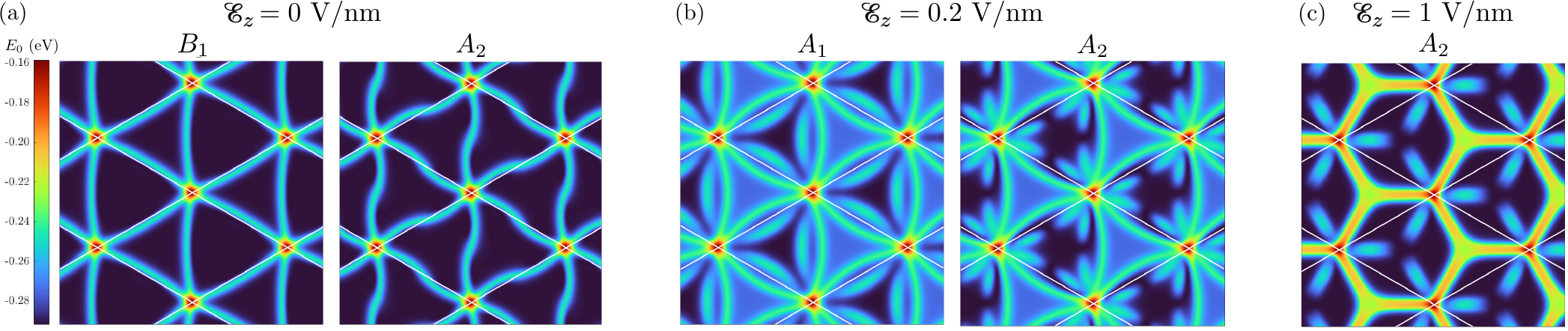}
    \caption{\textbf{Phonon distortion of the relaxation pattern at $\boldsymbol{\theta = 1^{\circ}}$}. (a) $\Gamma$-point phonon patterns at zero electric field. (b) With nonzero electric field. Note how the pattern of the $A_{1}$ distortion resembles a combination of the former $B_{1}$ breathing mode and the bending of the domain walls in the opposite direction due exclusively to relaxation (compare with Fig.~\ref{fig:fig1}(b) in the main text). The effect of the field on the $A_{2}$ moir\'e phonon can be understood from the inequivalent energy cost of fluctuations towards a MX or XM stacking region. (c) Honeycomb regime. There is only one singlet mode in the same range of energies.} 
\vspace{-0.5cm} 
\label{fig:S3}
\end{center}
\end{figure}

We consider now dynamical fluctuations of the stacking field in the background of the relaxed solution,
\begin{align}
    \vec{\phi}(\vec{r},t) = \vec{\phi}_{0}(\vec{r})+\delta\vec{\phi}(\vec{r},t).
\end{align}
In the harmonic approximation the fluctuation satisfies the following equations:
\begin{align}
    \rho\delta\ddot{\vec{\phi}} = 
    (\lambda+\mu)\vec{\nabla}\vec{\nabla}\delta \vec{\phi} + \mu\vec{\nabla}^{2}\delta \vec{\phi} 
    - 2\delta \phi_{j}\left.\frac{\partial \bar{\mathcal{F}}_{\textrm{adh}}}{\partial \phi_{j}\partial \vec{\phi}}\right|_{\vec{\phi}=\vec{\phi}_{0}}
     - 2\delta \phi_{j}\left.\frac{\partial \bar{\mathcal{F}}_{\textrm{elec}}}{\partial \phi_{j}\partial \vec{\phi}}\right|_{\vec{\phi}=\vec{\phi}_{0}}.
\end{align}
We introduce Fourier series as
\begin{align}
    \delta \vec{\phi}(t,\vec{r}) = \frac{1}{\sqrt{A}}\sum_{\vec{k}}\int\frac{d\omega}{2\pi}\delta\vec{\phi}(\omega,\vec{k})e^{i\vec{k}\vec{r}-i\omega t},
\end{align}
where $A$ is the total area of the system. The previous equations can be written as
\begin{align}\label{eqn:eigenvalue}
    \omega^{2}\rho \delta\vec{\phi}(\omega,\vec{k}) = (\lambda+\mu)\vec{k}\cdot(\vec{k}\cdot \delta\vec{\phi}(\omega,\vec{k})) + \mu \vec{k}^{2} \delta\vec{\phi}(\omega,\vec{k}) + 2\sum_{\vec{G}}\tilde{K}(\vec{G}) \cdot\delta\vec{\phi}(\omega,\vec{k}-\vec{G}),
\end{align}
where we have introduced a matrix with elements
\begin{align}
    \tilde{K}_{ij}(\vec{G}) = \frac{1}{A_{m}}\int d\vec{r}  \left( \left.\frac{\partial^{2}\bar{\mathcal{F}}_{\textrm{adh}}}{\partial \phi_{i}\partial\phi_{j}}\right|_{\vec{\phi}=\vec{\phi}_{0}}
    + \left.\frac{\partial^{2}\bar{\mathcal{F}}_{\textrm{elec}}}{\partial \phi_{i}\partial\phi_{j}}\right|_{\vec{\phi}=\vec{\phi}_{0}}
    \right) e^{-i\vec{G}\vec{r}},
\end{align}
with the integral taken over the moir\'e unit cell. Equation~\eqref{eqn:eigenvalue} can be read as an eigenvalue problem in the vector space defined by  
\begin{align}\label{eqn:eigensate}
    \delta\vec{\Phi}(\omega,\vec{q}) :=
    \begin{pmatrix}
        \delta\vec{\phi}(\omega,\vec{q})\\
        \delta\vec{\phi}(\omega,\vec{q}-\vec{G}_{1})\\
        \delta\vec{\phi}(\omega,\vec{q}-\vec{G}_{2})\\
        \vdots
    \end{pmatrix},
\end{align}
where $\vec{q}\in$ mBZ and $\vec{G}_{1},\vec{G}_{2},...$ are reciprocal vectors of the moir\'e lattice that we use in the numerical implementation. We have included the same number of shells as in the relaxation problem detailed above. We show in Fig.~\ref{fig:S3} the effect of different singlet modes on the energy landscape in real space, from zero field to the honeycomb regime.

\begin{figure}[t!]
\begin{center}
    \includegraphics[width=\columnwidth]{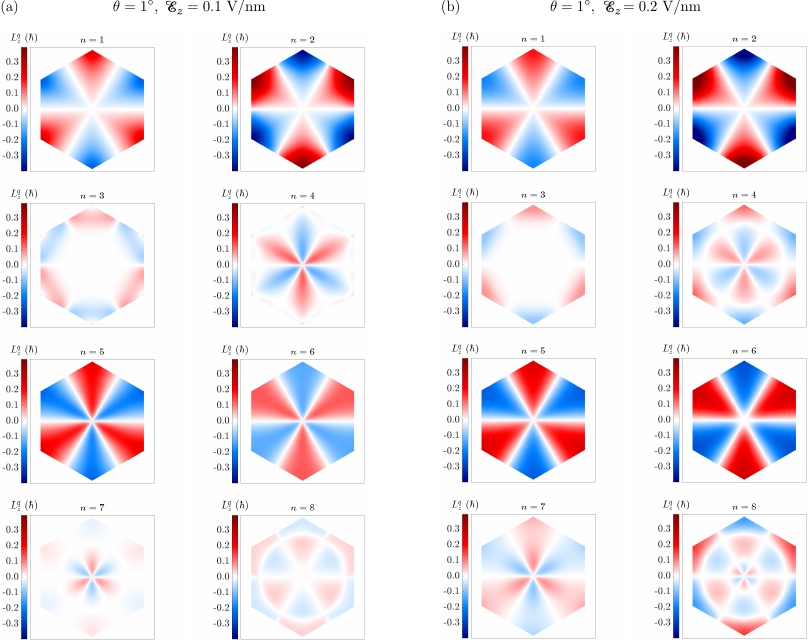}
    \caption{\textbf{Angular momentum of first eight bands of moir\'e phonons of MoS$_{2}$ within the mBZ}. The plots show the value of the angular momentum $L_{z}^{\vec{q},n}$ for a single twist angle $\theta = 1^{\circ}$. Each panel corresponds to a different electric field. Within each panel, $n$ is the band index from lower to higher energies (in particular, $n=1,2$ denote the longitudinal and transverse acoustic phasons, respectively).} 
\vspace{-0.5cm} 
\label{fig:S2}
\end{center}
\end{figure} 

\subsection{Vibrational density of states}

For each mode labeled by band index $n$ and momentum $\vec{q}$ we introduce a spectral function characteristic of a damped oscillator:
\begin{align}
    \mathcal{A}_n(\omega,\vec{q}) = \frac{2}{\pi} \frac{\omega_{n}^{2}(\vec{q})\gamma}{(\omega^{2} - \omega_{n}^{2}(\vec{q}))^{2}  +\gamma^{2}\omega^{2}}, 
\end{align}
where $\omega_{n}(\vec{q})$ is the mode frequency following from the previous diagonalization problem and $\gamma$ is a phenomenological damping coefficient. The vibrational density of states is the sum to all modes,
\begin{align}
    \mathcal{D}(\omega) = \sum_{n}\sum_{\vec{q}} \mathcal{A}_n(\omega,\vec{q}).
\end{align}
In Fig.~3 of the main text we took $\gamma/\omega_{\textrm{m}}=0.01$.

\subsection{Phonon angular momentum}

To compute the angular momentum of each band of moir\'e phonons we follow Ref.~\onlinecite{Koshino2020,Koshino2023}. For each $\vec{q}\in$mBZ, let $\delta\vec{\Phi}(\omega_{n},\vec{q})$ be the solution in the from of Eq.~\eqref{eqn:eigensate} of the eigenvalue problem in Eq.~\eqref{eqn:eigenvalue} corresponding to the $n-$th phonon band, normalized such that $\sum_{\vec{G}}|\delta\vec{\phi}(\omega_{n},\vec{q}+\vec{G})|^{2} = 1$. Then, in reciprocal space the angular momentum of the band is
\begin{align}
    L_{z}^{\vec{q},n} = i\hbar \sum_{\vec{G}} \delta\vec{\phi}(\omega_{n},\vec{q}+\vec{G}) \times \delta\vec{\phi}^{*}(\omega_{n},\vec{q}+\vec{G}).
\end{align}
A nonzero $L_{z}^{\vec{q},n}$ is only possible if $C_{2z}$ symmetry is broken, since $\delta\vec{\phi}(\omega_{n},\vec{q}+\vec{G})$ can be chosen to be real otherwise.

In Fig. \ref{fig:S2} we show the results for the first eight modes at twist angle $\theta=1^{\circ}$ and two values of the electric field, corresponding to the first two dispersions included in Fig. \ref{fig:fig1}(b) in the main text.
The difference in magnitude is explained by the change in energy of the modes at the K points and the relative growth in area of MM and MX/XM stacking regions \cite{xiao2021chiral}, although the former is the main contributor for the acoustic modes.
The field also affects the \text{spreading} of the angular momentum from the K points. The most significant example of this is seen in the fourth mode, which shows small regions of nonzero $L_{z}^{\vec{q},n}$ around the K points when $\mathcal{E}_{z}=0.1$ V/nm that grow after the field is increased to 0.2 V/nm. Finally, the qualitative details of the seventh and eighth modes are modified as a result of the band inversion (doublet and singlet) that occurs when increasing the field.

\end{document}